\documentclass{emulateapj}

\slugcomment{Accepted to ApJL}
\shorttitle{Disk-Halo Alignment}
\shortauthors{Bailin et al.}

\defcitealias{bjd98}{BJD98}
\defcitealias{kazantzidis-etal04}{K04}
\defcitealias{bs05-alignment}{BS05}

\newcommand{\rvir}{{\ensuremath{r_{\mathrm{vir}}}}}
\newcommand{\Mvir}{{\ensuremath{M_{\mathrm{vir}}}}}

\begin{document}

\title{Internal Alignment of the Halos of Disk Galaxies
 in Cosmological Hydrodynamic Simulations}

\author{Jeremy Bailin\altaffilmark{1,2}, Daisuke Kawata\altaffilmark{1,3},
Brad K. Gibson\altaffilmark{1}, Matthias
Steinmetz\altaffilmark{4}, Julio F. Navarro\altaffilmark{5},
Chris B. Brook\altaffilmark{6},
Stuart P. D. Gill\altaffilmark{1},
Rodrigo A. Ibata\altaffilmark{7},
Alexander Knebe\altaffilmark{1,4},
Geraint F. Lewis\altaffilmark{8},
Takashi Okamoto\altaffilmark{9,10}
}

\altaffiltext{1}{Centre for Astrophysics and Supercomputing,
 Swinburne University of Technology, PO Box 218, Hawthorn, Victoria,
 3122, Australia}
\altaffiltext{2}{jbailin@astro.swin.edu.au}
\altaffiltext{3}{Carnegie Observatories, 813 Santa Barbara St., Pasadena, CA,
  91101, USA}
\altaffiltext{4}{Astrophysikalisches Institut Potsdam, An der Sternwarte 16,
D-14482, Potsdam, Germany}
\altaffiltext{5}{Dept. of Physics \& Astronomy, University of Victoria, 
PO Box 3055 STN CSC, Victoria, BC, V8W 3P6, Canada}
\altaffiltext{6}{D{\'e}partement de physique, de g{\'e}nie physique et
 d'optique, Universit{\'e} Laval, Qu{\'e}bec, QC, G1K 7P4, Canada}
\altaffiltext{7}{Observatoire de Strasbourg, 11, rue de l'Universite, F-67000, Strasbourg, France}
\altaffiltext{8}{School of Physics, University of Sydney, NSW 2006, Australia}
\altaffiltext{9}{Institute for Computational Cosmology, Physics Department,
 Durham University, South Road, Durham, DH1 3LE, UK}
\altaffiltext{10}{National Astronomical Observatory of Japan, Mitaka, Tokyo
181-8588, Japan}

\begin{abstract}
Seven cosmological hydrodynamic simulations of disk galaxy formation are
analyzed to determine the alignment of the disk within the
dark matter halo and the internal structure of the halo.
We find that the orientation of the outer halo,
beyond $\sim 0.1~\rvir$,
is unaffected by the presence of the disk.
In contrast, the inner halo is aligned such that
the halo minor axis aligns with the disk axis. The
relative orientation of these two regions of the halo are uncorrelated.
The alignment of the disk and inner halo appears to take place
simultaneously through their joint evolution.
The disconnect between these two regions of the halo
should be taken into account when modelling tidal streams in
the halos of disk galaxies and when calculating intrinsic alignments
of disk galaxies based on the properties of dark matter halos.
\end{abstract}

\keywords{galaxies: formation --- galaxies: halos --- dark matter
--- galaxies: spiral --- galaxies: evolution}

\section{Introduction}
The alignment of galactic disks within their triaxial dark matter halos is both
observationally and theoretically difficult to determine. Observationally,
dark matter can only be traced by its gravitational effect on luminous matter.
Efforts to determine the shapes of individual dark matter halos around disk
galaxies usually rely on the motions of tidal streams
\citep{johnston-etal99,jlm05,ibata-etal01,ibata-etal04,helmi04a,helmi04b,martinez-delgado-etal04,ljm05},
polar rings \citep{sackett-etal94},
or the flaring of the gas layer \citep{om00}.
Most of these studies have found that the halo is flattened along
the pole of the galactic disk; however, the degree of flattening for even
the best-studied case of the Milky Way is controversial, with values
ranging from 0.8 to 1.7 (i.e. \textit{elongated} along the Galactic pole).

Theoretical determinations of galactic disk alignment
within halos have been restricted
by available computing power. $N$-body simulations have been
used to determine the flattening and internal alignment
of pure dark matter halos (\citealp{be87,warren-etal92,js02};
\citealp{bs05-alignment}, hereafter \citetalias{bs05-alignment}).
These studies have found that the axes of halos, especially
the minor axes, are very well aligned internally; the halo is
well approximated by a set of concentric ellipsoids.
Under the assumption that the angular momentum of the baryons which form
the disk is aligned with the angular momentum of the dark matter,
the disk axis is expected to lie typically 25\degr\ from the halo minor
axis \citepalias{bs05-alignment}. However, hydrodynamical simulations suggest
that the angular momentum of the baryons and dark matter are not
perfectly aligned \citep{vdb-etal02,ss05}. Moreover, the 
orientation of the halo may change in the presence of a misaligned disk 
(\citealt{bjd98}, hereafter \citetalias{bjd98}).

This problem can now be addressed with recent high resolution simulations
of disk galaxy formation.
These simulations have traditionally formed disks that are
much more compact than observed galactic disks \citep{ns97}.
However, modern treatments of feedback
and more consistent analyses of the simulations and observations
have led to much better agreement:
the physical size of the disks is realistically reproduced, and the lack
of angular momentum manifests itself primarily in a very
pronounced slowly rotating bulge-halo component \citep{abadi-etal03a}.
Recent works
employing advanced multi-phase descriptions of the ISM reduce the
dominance of the bulge, in particular if feedback from AGNs is
included \citep{robertson-etal04,okamoto-etal05}. Since the physical
size of the disks is realistically reproduced and since
the location of the surrounding matter which provides
the tidal torque is unaffected,
the orientation of the angular
momentum, and therefore the disk, is expected to be robust.

\citet{nas04} found that in four such simulations, the disk axis
tends to lie within 30\degr\ of the intermediate axis of the large scale
mass distribution at turnaround,
indicating that primordial tidal torques play an important
role in determining the final disk orientation.
It is known that the radial distribution of dark matter in the halo
changes in the presence of cooled baryons \citep{gnedin-etal04},
and therefore it is likely that other properties of the matter
distribution are affected too.
\citet{kazantzidis-etal04} (hereafter 
\citetalias{kazantzidis-etal04}) studied the change in the axes
of the halo of a high resolution disk galaxy simulation compared to the
$N$-body case.
They found that the axis ratios of the halo
were dramatically higher (i.e. more spherical)
when the simulation included baryon cooling than when the baryons were
not present or not allowed to cool to a disk. They also found that
the minor axis of the halo was aligned with the disk axis
within $r<0.2~r_{180}$,
but that the halo became \textit{elongated} along the disk axis beyond
that radius.

The alignment of disks with their dark matter halos is important for
understanding a number of unresolved questions regarding disk galaxy
formation, evolution, and dynamics. Warps in disk galaxies are ubiquitous
\citep{rc98,sd01,gsk02} and have been proposed to be caused by misalignment
between the disk and the dark matter halo (\citealp{ds83,toomre83,bailin-phd};
see however \citealt{nt95}; \citetalias{bjd98}). This possibility can be ruled out if
such misalignments never occur in fully self-consistent simulations of disk
galaxy formation. The Holmberg effect 
(\citealp{holmberg-effect,zaritsky-etal97}; see however
\citealp{brainerd04,willman-etal04})
may be explained if galactic disks preferentially lie perpendicular to the
major axis of their dark matter halo \citep{knebe-etal04,zentner-etal05}.
Models of the Sagittarius tidal stream cannot simultaneously fit the leading
and trailing arm of the stream with the same halo flattening 
\citep{helmi04b,ljm05};
a more detailed understanding of the relationship between the
location of the disk and the
shape of the dark matter halo may be necessary to resolve this issue.
Finally, the intrinsic alignment of observed galaxies acts as a background
contaminant in weak lensing studies. Predictions of the magnitude of this
effect based on linear theory or
$N$-body simulations must assume a disk orientation for
each halo, which is as yet untested \citep{hrh00,pls00,cnpt01}.

In this Letter, we analyze the shape and internal alignment of the dark
matter halos of seven high resolution cosmological disk galaxy formation
simulations, and
discuss the possible observational effects of such internal structure.

\section{Methodology}

We analyze seven cosmological simulations which use the multi-mass technique
to self-consistently model the large scale tidal field while simulating
the galactic disk at high resolution. These simulations 
include self-consistently almost all the important physical processes
in galaxy formation, such as self-gravity, hydrodynamics,
radiative cooling, star formation, supernovae feedback and
metal enrichment.

\begin{deluxetable}{llrrr}
\tablewidth{0pt}
\tablecaption{%
Properties of simulations\label{simulation table}}
\tablehead{ \colhead{Simulation Name} & \colhead{Code} &
  \colhead{\Mvir} & \colhead{\rvir} & \colhead{$r_{\mathrm{disk}}$} \\
 & & \colhead{($M_{\sun}$)} & \colhead{(kpc)} & \colhead{(kpc)} }
\startdata
KGCD & GCD+ & $8.8 \times 10^{11}$ & 240 & 10\\
AGCD & GCD+ & $9.3 \times 10^{11}$ & 270 & 21\\
AGSPH & GRAPESPH & $9.4 \times 10^{11}$ & 270 & 23\\
KIA5 & GRAPESPH & $1.7 \times 10^{12}$ & 300 & 4\\
KIA9 & GRAPESPH & $2.1 \times 10^{12}$ & 320 & 20\\
KIB1 & GRAPESPH & $2.9 \times 10^{12}$ & 390 & 16\\
KIB2 & GRAPESPH & $8.5 \times 10^{11}$ & 260 & 5\\
\enddata
\end{deluxetable}

The properties of the simulations are listed in Table~\ref{simulation table}.
Column 1 contains the name by which we refer to each simulation,
Column 2 is the code used to run the simulation, Column 3 is the virial
mass of the galaxy, Column 4 is the virial radius of the galaxy,
and Column 5 contains the radial extent of the gas disk in each
simulation, defined as the largest radius at which we find gas particles
in the disk plane.
The primary simulation we analyze is the ``KGCD'' simulation, performed
using the GCD+ code \citep{kg03a}. The simulation is a higher
resolution model of galaxy ``D1'' in \citet{kgw04}. 
The mass and softening length of 
individual gas (dark matter) particles 
in the highest-resolution region are $9.2\times10^5$ 
($6.2\times10^6$) ${\rm M}_\odot$ and 0.57 (1.1) kpc, respectively.
We analyze the KGCD simulation both at the final output at
$z=0.10$ and an output at $z=0.37$ (approximately 2.2~Gyr earlier) at which
the disk first appears fully formed in approximately its final state
to determine whether there is significant evolution in the halo
structure in the absence of major mergers.
In order to ensure that our results are not particular to this
code, the feedback prescription, or the particular set of initial
conditions, we also analyze the
final snapshot of the simulation described in 
\citet{abadi-etal03a} (``AGSPH''), a simulation with the same 
initial conditions as AGSPH but run using GCD+ (``AGCD''),
and four additional GRAPESPH \citep{grapesph}
simulations: KIA5, KIA9, KIB1, and KIB2.
Dark-matter-only versions of each of these simulations were also run
in order to directly compare the
alignment of each halo with and without a galactic disk.

The principal axes and axis ratios of the dark matter halo are calculated
using the method of \citetalias{bs05-alignment}. Briefly, the particles
are split into spherical shells bounded by radii of
$r = 0.025, 0.05, 0.1, 0.25, 0.5,$ and $1.0~\rvir$%
\footnote{We use the fitting function from
Appendix~A of \citet{ks96} to calculate
the overdensity that defines the virial radius \rvir\ as a function
of cosmology and $z$}.
The reduced inertia tensor is calculated for the particles within
each shell and diagonalized to the determine the principal axes.
Finally, the axis ratios determined from the eigenvalues of
the reduced inertia tensor are corrected for the bias introduced by
using spherical shells.
In this way, the results at each radius are entirely independent
\citepalias{kazantzidis-etal04}.

\section{Results}%
\label{results section}

\begin{figure}
\plotone{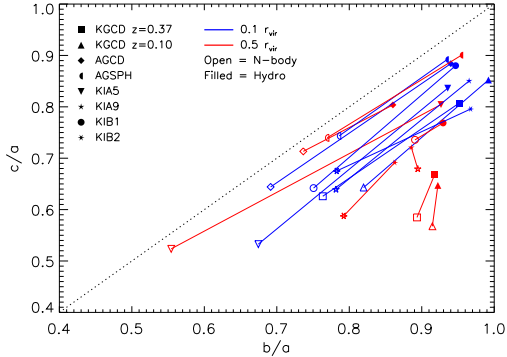}
\caption{\label{axis ratio plot}%
Axis ratios $b/a$ and $c/a$ of the dark matter halos in simulations
including baryonic physics compared to simulations containing only
dark matter.
Blue symbols and lines
(dashed liens)
represent the axes measured at $0.1~\rvir$,
while red symbols and lines
(solid lines)
represent the axes measured
at $0.5~\rvir$. Filled symbols represent the simulations containing
baryonic disks, while open symbols represent the $N$-body only
simulations. Lines connect simulations with the same initial
conditions and show the change in axis ratios induced by the
inclusion of baryonic physics. The different symbols refer to
the different simulations.%
}
\end{figure}

The change in axis
ratio for our simulations is shown in Figure~\ref{axis ratio plot}.
An increase in sphericity is clearly evident in this figure, with a magnitude
that decreases at larger radii.
The magnitude of
this change is smaller than that found by \citetalias{kazantzidis-etal04},
who also found that the presence of baryonic
cooling dramatically increases the axis ratios of dark matter halos,
with changes in $b/a$ and $c/a$ that range from $0.35$ at small radii
to zero at the virial radius. The change in axis ratios was particularly
evident in their galactic disk simulation, where $b/a$ rose by
almost $0.6$ when a galactic disk was present. 
This difference in magnitude
may simply be small number statistics: the $N$-body version of the
galactic disk simulation of \citetalias{kazantzidis-etal04} is exceptionally
flattened, and therefore a much greater increase in axis ratio is
possible.

\begin{figure}
\plotone{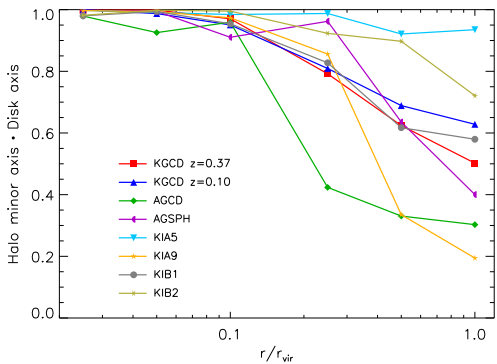}
\caption{\label{alignment plot}%
The alignment between the minor axis of the dark matter halo
at different radii and the disk axis. The different colors and symbols
represent different simulations.%
}
\end{figure}

\begin{figure}
\plotone{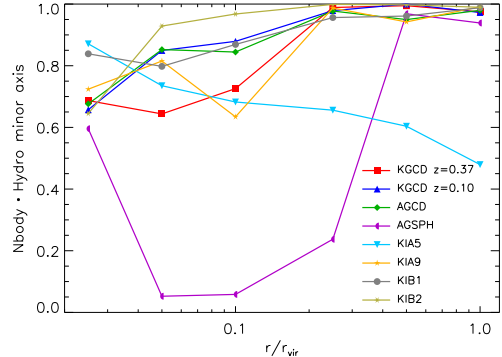}
\caption{\label{nbody alignment plot}%
The alignment between the minor axis of the dark
matter halo of each hydrodynamic simulation and of the
matching $N$-body simulation, as a function of radius
within the halo. The different colors and symbols represent
different simulations.%
}
\end{figure}

The direction cosine between the disk axis and the halo minor axis
at each radius is shown
in Figure~\ref{alignment plot},
while Figure~\ref{nbody alignment plot}
shows the change in the alignment of the minor axis at each radius
compared to the dark-matter-only version of the same simulation.
This is unity if the axes are perfectly aligned and vanishes if they
are perpendicular.
The disk axis is defined to be the minor axis of the gas distribution
within $r_{\mathrm{disk}}$;
in all cases, this matches the axis determined by
visual inspection, and that determined including all baryons
within $r_{\mathrm{disk}}$.
Two regimes are clearly visible in Figures~\ref{alignment plot}
and~\ref{nbody alignment plot}:
\begin{enumerate}
 \item $r < 0.1~\rvir$: In the inner regions of the halo, the minor axis
	is extremely well aligned with the disk axis.
 \item $r > 0.1~\rvir$: In the outer regions of the halo, the orientation
	of the minor axis is unchanged from the $N$-body only case.
\end{enumerate}
The transition between these regimes occurs at a slightly larger
radius in the AGSPH and KIA5 simulations;
as these simulations contain respectively the largest and smallest disks,
the transition radius does not appear to depend on the disk radius.

KIA5 appears as an outlier on all of these plots: it has a large change
in axis ratio even at large radius, it shows very little change in
orientation between the inner and outer halo, and the minor axis in the
hydrodynamic simulation is not at all
aligned with the $N$-body case.
This latter point may be because the $N$-body version of
KIA5 is exceptionally prolate in its outer regions, making a determination
of the minor axis ill-defined.
However, we note that KIA5 contains the smallest
gaseous disk and may represent an earlier-type galaxy than the
other simulations, with a disk that is much less important to its
dynamics. This interpretation is bolstered by the fact that KIB2,
which contains the second-smallest disk after KIA5, also shows
smaller than average misalignment between its inner and outer regions.

The distribution of cosines between randomly-oriented unit vectors is
uniform with a mean of 0.5. This is precisely the distribution of cosines
between the inner and outer halo axes in Figure~\ref{alignment plot}.
Therefore, we conclude that the orientation of the inner and outer halo
in simulations containing galactic disks are uncorrelated.
In contrast, the minor axes of pure dark matter halos are very well
aligned throughout their entire extent, with
a direction cosine of always greater than $0.88$ (see Figure~8b of
\citetalias{bs05-alignment}).
Therefore, the presence of the disk strongly modifies the
shape of the inner halo, reorienting it so that the halo minor axis aligns
with the disk axis.
This is in qualitative agreement with
\citetalias{bjd98}, who discussed the effect of misalignment on
the disk rather than the halo, but concluded that the halo reorients
itself to line up with the disk axis.
The results for the KGCD simulation at $z=0.10$ and $z=0.37$ are very
similar, indicating that either this reorientation must happen
on a short timescale, or that the processes that determine the disk
and halo axes act on both simultaneously during their formation.

There is some relation between the disk axis and the minor axis of
the unperturbed halo, as the distribution of innermost points
in Figure~\ref{nbody alignment plot} has a mean greater than 0.5;
this is expected due to the tendency of the halo angular momentum to
align with its minor axis (\citealp{warren-etal92};
\citetalias{bs05-alignment}).
However, this correlation is not strong and the slight misalignment
within pure dark matter halos ensures that there is no residual
correlation between the inner and outer regions of halos containing
disks.

\section{Discussion}

The final orientation of the galactic disk in a cosmological dark
matter halo is still unresolved. It is clearly related to the angular
momentum of the material in the halo, and the results of \citet{nas04}
indicate that the primordial tidal torques of the large scale
matter distribution play an important role, but there is considerable
galaxy-to-galaxy variation in the results of the simulations. While
the primordial tidal torques generate the angular momentum in the material
which eventually becomes the disk, the accretion of this material
by the disk
is a clumpy, stochastic process \citep{vitvitska-etal02}, and therefore
we are not able to predict the precise orientation of the disk.

However, the results presented in \S~\ref{results section}
clearly indicate that
the halo orientation within $0.1~\rvir$
matches the disk axis.
Beyond $0.1~\rvir$, the halo orientation is unaffected by the presence
of the disk, although the axis ratios are somewhat larger 
\citepalias{kazantzidis-etal04}.
This outer region is itself internally well-aligned, resulting in two
distinct regions of the halo with unique and uncorrelated orientations.
These results are independent of the time that the halo is studied,
the code used, and the particular galaxy simulated.

It is interesting to consider whether the disk drives the orientation of
the inner halo or vice versa. While the disk is less
massive than the inner halo (the baryons 
represent $15$ -- $42\%$ of the mass within
$0.1~\rvir$), it is also much more flattened than the halo.
To evaluate which is most important,
we have constructed rings of test particles tilted by
$10\degr$ with respect to the disk plane and
compared the gravitational torque on these test particles
due to the dark matter with that due
to the baryons. For rings of radius less than $\sim 0.1~\rvir$,
i.e. in the aligned region,
the torques from the dark matter and baryons are of comparable
magnitude, while the dark halo dominates the torque at larger radii.
Therefore, it appears that the alignment is due to the simultaneous
evolution of the disk and halo rather than the disk directly
driving the halo orientation or vice versa.

If the structure of the stellar halo is determined by the dark halo,
we may expect the properties of the stellar halo to change abruptly
at $0.1~\rvir$. In fact, the flattening of the stellar halo
of the Milky Way
changes from a flattened distribution aligned with the disk
at $R<15~\mathrm{kpc}$ to an essentially spherical distribution
at $R\approx 20~\mathrm{kpc}$ \citep{cb00}. This spherical distribution
may be an intermediate stage between
two misaligned flattened regions.
However, the virial radius of the Milky Way is
likely at least $250$~kpc, indicating that this observed transition
occurs at a smaller radius than the transition seen in our
simulated halos.

Because the disk is always aligned with the halo in the inner regions,
general misalignment between the disk and halo is not a viable mechanism for
generating galactic warps. 
However, if very low surface density disk material existed beyond
$0.1~\rvir$,
it may settle into the outer halo symmetry plane, and
therefore be warped with respect to the inner disk. Indeed,
\citetalias{bjd98} suggested that the warp modes within live halos
would be essentially different from those in static halos not in
the configuration of the disk, but rather in the configuration of
the halo.
However, the outer extent of the baryonic disk is determined by the
material which has most recently cooled; this material is essentially
unaffected by the angular momentum problem \citep{ns97}.
Therefore, the maximum extent of simulated disks
is expected to be accurate, and therefore the disk
is unlikely to extend into the misaligned outer halo.
However, simulations of such systems should be carried out to determine
if such warped configurations are stable, and to determine
how the surface density of the disk affects the tendency of the
halo to realign itself.

The solution to the Holmberg effect proposed by \citet{knebe-etal04}
requires that galactic disks are preferentially perpendicular to the
major axis of the halo. However, we find that the orientation of the disk
is uncorrelated with the orientation of the major axis of the
outer halo, where most satellite galaxies lie, casting
doubt on this hypothesis.

Finally, we note that the orbit of the Sgr dSph in the Milky Way
contains segments both inside and outside of 0.1~\rvir\ \citep[e.g.][]{hw01}.
If the orientation of the outer Milky Way halo is significantly different
from the orientation of the Galactic disk, as suggested by our
results, then the orbits of particles in the Sagittarius stream may differ
from those calculated assuming that the halo can be approximated by
concentric ellipsoids
(note, however, that the stream depends on the
flattening and orientation of isopotential contours,
which are generated by matter at a range of radii and therefore
change more smoothly than the density).
This additional ingredient may be required
to self-consistently model both the leading and trailing arms
of the stream.

\acknowledgements
JB thanks Chris Power for useful discussions. DK acknowledges the
Japan Society for the Promotion of Science
postdoctoral fellowship for Research Abroad.
TO acknowledges the Japan Society for the Promotion of Science
for Young Scientists (No. 1891).

\bibliography{ms.bib}

\end{document}